\documentclass[journal=jacsat,manuscript=article]{achemso}

\usepackage[version=3]{mhchem} 
\newcommand{\vcr}[1]{{\bf #1}}
\newcommand{\vc}[1]{\vec{#1}}
\usepackage{amssymb}
\usepackage{color}
\newcommand{\rmm}[1]{{\textcolor{black}{#1}}}

\author{Ra\'i M. Menezes}
\affiliation{NANOlab Center of Excellence \& Department of Physics, University of Antwerp, Groenenborgerlaan 171, B-2020 Antwerp, Belgium}

\author{Jeroen Mulkers}
\affiliation{DyNaMat Lab, Department of Solid State Sciences, Ghent University, Ghent, Belgium}

\author{Cl\'ecio C. de Souza Silva}
\affiliation{Departamento de F\'isica, Universidade Federal de Pernambuco, Cidade Universit\'aria, 50670-901, Recife-PE, Brazil}

\author{Bartel Van Waeyenberge}
\affiliation{DyNaMat Lab, Department of Solid State Sciences, Ghent University, Ghent, Belgium}

\author{Milorad V. Milo\v{s}evi\'c}
\email{milorad.milosevic@uantwerpen.be}
\affiliation{NANOlab Center of Excellence \& Department of Physics, University of Antwerp, Groenenborgerlaan 171, B-2020 Antwerp, Belgium}
\title[An \textsf{achemso} demo]
  {Towards Magnonic Logic and Neuromorphic Computing: Controlling Spin-Waves by Spin-Polarized Current}


\begin{document}




\begin{abstract}
  Spin-waves (magnons) are among the prime candidates for building fast yet energy-efficient platforms for information transport and computing. We here demonstrate theoretically and in state-of-the-art micromagnetic simulation the effects that strategically-injected spin-polarized current can have on controlling magnonic transport. We reveal analytically that the Zhang-Li spin-transfer-torque induced by applied current is analogous to the Dzyaloshinskii-Moriya interaction for scattering the magnons in the linear regime, to then provide a generalized Snell's law that describes the spin-wave propagation across regions with different current densities. We validate the latter in numerical simulations of realistic systems, and exemplify how these findings may help advance the design of spin-wave logic and neuromorphic computing devices.
\end{abstract}


The development of neuromorphic computing hardware has attracted significant attention in recent years as such platforms are capable of performing complex information processing tasks, such as classification and pattern recognition of various types of data, from e-commerce to scientific content~\cite{christensen20222022,romera2018vowel,shainline2017superconducting}. A central challenge of this research is the requirement of highly interconnected systems, inspired by the biological concepts of the human brain. Interestingly, wave-based physical systems have been demonstrated to operate as recurrent neural networks~\cite{hughes2019wave}, where interference patterns in the propagating substrate can realize an all-to-all interconnection between points of the substrate that mimic the action of artificial neurons by scattering and recombining input waves in order to extract their information. 

Similar to other wave phenomena in physics, spin-waves (magnons) travel through space accompanied by a transfer of energy, which if precisely controlled can lead to fast information transport and computing applications in the nanometric to micrometric scale~\cite{mahmoud2020introduction,barman20212021}. Spin-waves are readily demonstrated as a promising platform for performing logic operations~\cite{chumak2014magnon,chumak2015magnon} and the recent theoretical advances in wave-based computation can pave the way for spintronic hardware in the field of artificial intelligence~\cite{papp2021nanoscale}. However, even though extensive research has been carried out in recent years, the precise manipulation of spin waves in nanostructures has not been entirely mastered and needs to be advanced for the benefit of functional magnonic devices. 

One manner of manipulating the spin textures in magnetic materials is by the application of spin-polarized (SP) currents, having (part of the) spins of the moving electrons aligned. The interaction of a SP current with the localized magnetic moments results in a torque on the magnetization, dubbed a \rmm{Zhang-Li} spin-transfer-torque (STT)~\cite{Zhang2004}. Being able to affect the orientation of the magnetization, STT has become nearly unavoidable in the design of spintronic nanodevices~\cite{Ralph2008}. The effect and applicability of the STT was demonstrated in racetrack memory concepts, where the position of local magnetic structures, such as domain walls and skyrmions, is controlled by in-plane currents~\cite{wang2022electrical,peng2021dynamic}. Moreover, not only do SP currents modify the equilibrium magnetic state, they can also significantly influence the propagation of spin waves (SWs). For instance, in Ref.~\cite{Bazaliy1997} the authors derived the dispersion relation of SWs in a ferromagnet subjected to a uniform SP current. The applied current introduces a term proportional to the wavevector $k$ in the spin-wave dispersion relation. The nonreciprocity in the dispersion relation for waves with a $k$ vector in the same or the opposite direction as the current flow causes Doppler shift~\cite{fernandez2004influence}, as first validated experimentally by Vlaminck and Bailleul~\cite{Vlaminck2008}.

\begin{figure}[!t]
\centering
\includegraphics[width=0.6\linewidth]{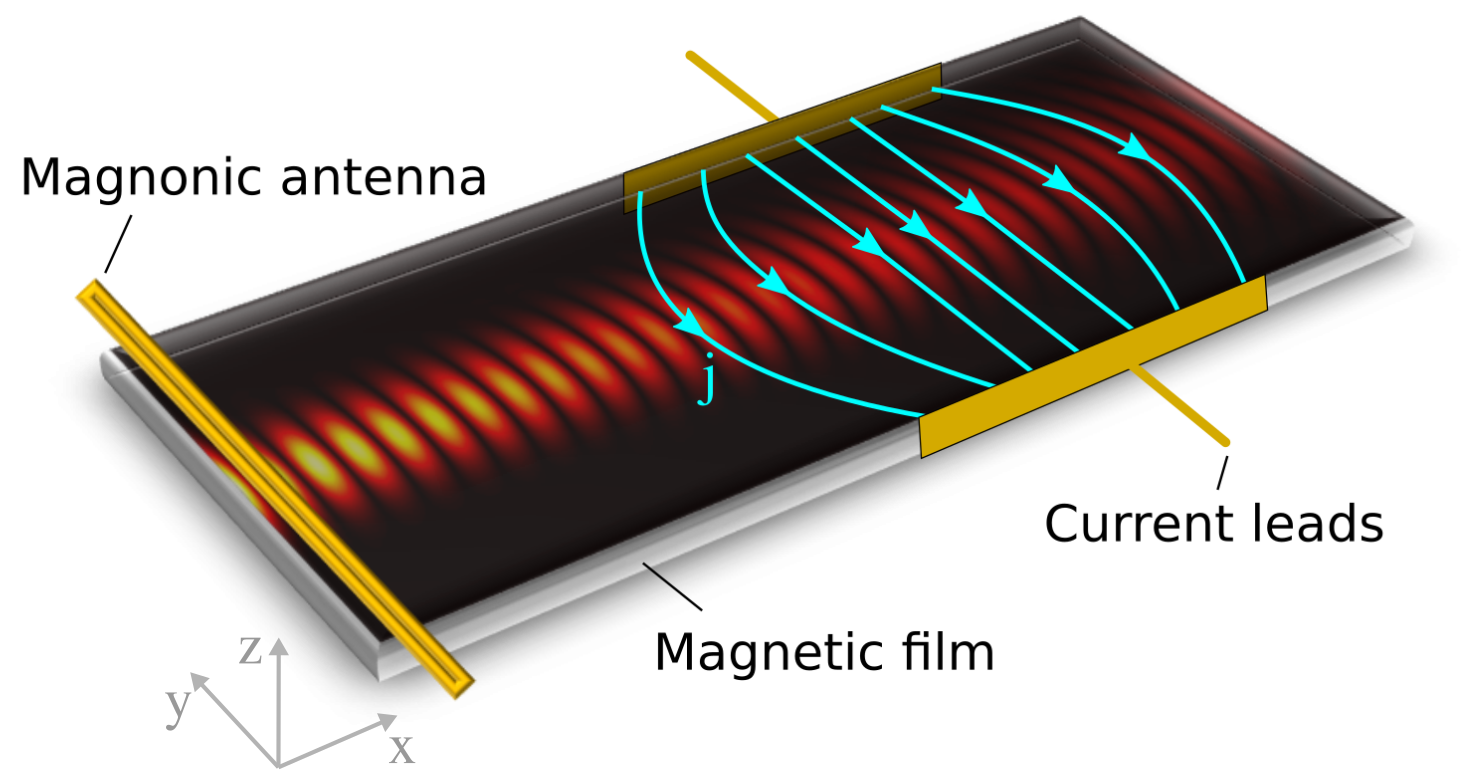}
\caption{\textbf{Schematic illustration of a SW facing a non-uniform distribution of the SP current}. The SW (in red) is induced by the input antenna on the left side and propagates along the magnetic film. Voltage electrodes induce the distributed SP current ($j$) which can be tuned to control the SW propagation.}
\label{fig1}
\end{figure}

A similar nonreciprocal term in the dispersion relation can also be induced by an electric field~\cite{Mills2008} or by the antisymmetric exchange interaction, also known as Dzyaloshinskii-Moriya interaction (DMI)~\cite{Moon2013}. Analogously to the Doppler shift induced by the currents, a frequency shift has been measured in ferromagnetic materials possessing DMI~\cite{cortes2013influence}. For heterochiral ferromagnets (\textit{i.e.}, with spatially varying DMI), the reciprocal term causes a nontrivial refraction of spin waves at interfaces between regions with different DMI, as described by a generalized Snell's law in Ref.~\citenum{mulkers2018tunable}. In this regard, the equivalency in the dispersion relations of SWs in the presence of DMI and SP current suggests that a spatially varying current density (as illustrated in Fig.~\ref{fig1}) can be used to manipulate the propagation of SWs.

In this Letter, we detail the effect of a SP current on controlling the propagation of spin-waves. We show that the \rmm{Zhang-Li} STT induced by \rmm{in-plane} current has analogous effect to DMI for confining and controlling the propagation direction of magnons in the linear regime. We proceed to derive a Snell's law to describe the scattering of spin-waves between regions with different current densities, and validate it by advanced simulations including solving Poisson's equation within the micromagnetic framework. Finally we present selected tailored examples to illustrate how strategically applied current can be employed to advance logic and neuromorphic computing devices based on spin-waves.

\section{Results and discussion}

\subsection{Spin-wave dispersion under applied current}

Within the micromagnetic framework, we describe the magnetization of a thin extended ferromagnetic film by considering the vector field $\vc{M}(\vcr{r}) = M_{\text{s}} \vc{m}(\vcr{r})$ with constant magnetization modulus $|\vc{M}|= M_{\text{s}}$ and the normalized magnetization direction $\vc{m}(\vcr{r})$ at each point $\vcr{r} \in \mathbb{R}^2$ of the film \footnote{In this paper we use an arrow for 3D vector fields such as the magnetization $\vc{m}$ and the effective magnetic field $\vc{H}_{\text{eff}}$ and bold variables for 2D dimensional vectors in the plane such as the position $\vcr{r}$, the $\vcr{k}$-vector, and the current~$\vcr{u}$.}. The dynamics of the magnetization is governed by the Landau-Lifshitz-Gilbert (LLG) equation
\begin{equation}\label{eq.llg}
    \dot{\vc{m}} = - \gamma \vc{m} \times \vc{H}_{\text{eff}} + \alpha \vc{m}\times\dot{\vc{m}} + \vc{\tau}_{\text{STT}},
\end{equation}
where $\gamma$ is the gyromagnetic ratio, $\alpha$ the dimensionless damping factor, and $\vc{H}_{\text{eff}}$ the effective field, which can be derived from the free energy~$E[\vc{m}]$ by taking the functional derivative with respect to the magnetization: $\vc{H}_{\text{eff}} = - \delta E/\delta \vc{M} $.

We extend the LLG equation by adding the torque~$\vc{\tau}_{\text{STT}}$ which includes the adiabatic and non-adiabatic STT terms derived by Zhang and Li~\cite{Zhang2004}:
\begin{equation}\label{eq.stt}
    \vc{\tau}_{\text{STT}} = - \vc{m}\times(\vc{m}\times(\vcr{u}\cdot \nabla)\vc{m}) + \beta \vc{m}\times(\vcr{u}\cdot \nabla) \vc{m},
\end{equation}
where
\begin{equation}\label{eq.u}
    \vcr{u} = -\frac{\mu_B P}{e M_{\text{s}} (1+\beta^2)}\vcr{j}.
\end{equation}
Here, $\nabla$ is the two-dimensional differential operator, $\beta$ is a dimensionless constant that represents the degree of non-adiabaticity, $e$ the elementary charge and $\mu_B$ the Bohr magneton. The polarization $P$ is a property of the ferromagnet and does not depend on the magnetization. Although $\vcr{u}$ has the units of velocity, we refer to it as \emph{current} since it is proportional to the SP current density~$\vcr{j}$. The most prominent effect of the STT is best understood by assuming small dissipation terms ($\alpha\approx 0$ and $\beta\approx 0$). In that case it is easy to prove that the solution of the LLG equation is given by $\vc{m}(\vcr{r}-\vcr{u}t;t)$ if $\vc{m}(\vcr{r};t)$ is the solution in absence of the STT. Put differently, adding the STT shifts the solution with a velocity $\vcr{u}$ \cite{Thiaville2005}. For instance, due to the STT, relaxed local structures such as domain walls and skyrmions will move with a velocity~$\vcr{v}=\vcr{u}$ when the current is switched on. Accordingly, a SW packet traveling in the film will gain an additional velocity equal to $\vcr{u}$. This insight will help to obtain an intuitive understanding of the results presented in this paper in which we study the effect of a non-uniform static current~$\vcr{u}(\vcr{r})$ on the propagation of SWs.

The dynamics of the magnetization depends strongly on the characteristics of the magnetic film, incorporated in the free energy functional. In this paper, we take into account the contribution of exchange interaction and Zeeman energy due to applied bias magnetic field:
\begin{equation}\label{eq.E}
    E[\vc{m}] = \iint \left[ A(\nabla\vc{m})^2 - \vc{H} \cdot \vc{m}\right] \text{d} x \text{d} y,
\end{equation}
with exchange stiffness $A>0$ and bias field $\vc{H}$.

The ground state magnetization is the ferromagnetic (field-polarized) state in which the magnetization is aligned with the bias field $\vc{H}$. To study small deviations from the ground state it is useful to construct the right-handed coordinate system $(\hat{e}_a,\hat{e}_b,\hat{e}_0)$ with $\hat{e}_0 \parallel \vc{H}$. The effective field in this coordinate system becomes
\begin{equation}\label{eq.Heff}
    \vc{H}_{\text{eff}}(\vcr{r},t) =
    \begin{bmatrix}
        \frac{2A}{M_{\text{s}}} \nabla^2 m_a(\vcr{r},t) \\
        \frac{2A}{M_{\text{s}}} \nabla^2 m_b(\vcr{r},t) \\
        \frac{2A}{M_{\text{s}}} \nabla^2 m_0(\vcr{r},t)+H_0
    \end{bmatrix},
\end{equation}
where $\vc{m} = (m_a,m_b,m_0)$ and $H=(0,0,H_0)$. Here and in the next section, we focus on first order deviations from the ground state and consider the SW solution as $m_0\approx1$, $m_a=A_0e^{i(\textbf{k}\cdot\textbf{r}-\omega t) - \mu t}$ and $m_b=iA_0e^{i(\textbf{k}\cdot\textbf{r}-\omega t) - \mu t}$, where $A_0\ll1$ represents the SW amplitude; $\omega$ is the SW angular frequency; $\textbf{k}$ is the wave vector, and $\mu$ represents the damping of the SW. Substituting that into the LLG equation [Eq.~\eqref{eq.llg}], with STT term given by Eq.~\eqref{eq.stt} and effective field from Eq.~\eqref{eq.Heff}, yields the dispersion relation (see supplementary material~\cite{SM})
\begin{equation}
    \omega = \frac{\gamma H_0}{1+\alpha^2}\left(1+\xi^2 k^2\right)+\frac{1+\alpha\beta}{1+\alpha^2} \vcr{u} \cdot \vcr{k},
\end{equation}
and damping parameter
\begin{equation}\label{eq.mu}
    \mu = \alpha \omega - \beta \vcr{u}\cdot \vcr{k},
\end{equation}
where we define the length scale $\xi=\sqrt{2A/H_0M_s}$. Note that the damping $\mu$ of the SW decreases if the wave vector~$\vcr{k}$ is parallel to the electron flow~$\vcr{u}$. Consequently, the attenuation length of a spin wave can be increased by applying a current opposite to the propagation direction, as predicted earlier by Seo \emph{et al.}~\cite{Seo2009}.

\textit{\textbf{Propagation direction}.}---The dispersion relation consists out of circular isofrequencies in $k$-space, which are shifted away from the origin due to the current. The center of the circular isofrequency is given by
\begin{equation}
    \vcr{k}_0 = - \frac{1+\alpha\beta}{2\gamma H_0 \xi^2} \vcr{u},
\end{equation}
and the radius by
\begin{equation}
    k_g = \left| \vcr{k}-\vcr{k}_0 \right| = \sqrt{\frac{1+\alpha^2}{\gamma H_0 \xi^2}(\omega - \omega_0)},
\end{equation}
with the minimal frequency $\omega_0 = \gamma H_0(1-\xi^2k^2_0)$. 

The propagation velocity $\textbf{v}$ of a wave packet is given by the gradient of the frequency in $k$-space:
\begin{equation}
    \textbf{v} = \nabla_k \omega = \frac{2\gamma H_0 \xi^2}{1+\alpha^2}\vcr{k} + \frac{1+\alpha\beta}{1+\alpha^2}\vcr{u} =\frac{2\gamma H_0 \xi^2}{1+\alpha^2}(\textbf{k}-\textbf{k}_0),
\end{equation}
where $\textbf{k}_g=\vcr{k}-\vcr{k}_0$ defines the propagation direction. The velocity of the wave packet can be separated as 
\begin{equation}
    v_{\parallel} = \frac{\gamma H_0 \xi^2}{1+\alpha^2}2k + \frac{1+\alpha\beta}{1+\alpha^2}u_{\parallel},~
    v_{\perp} =\frac{1+\alpha\beta}{1+\alpha^2}u_{\perp},
\label{eq.velo}
\end{equation}
where $\parallel$ and $\perp$ denote the component parallel and perpendicular to the wave vector respectively. Notice that, in general, the propagation direction is not parallel to the wave vector $\textbf{k}$, and the SW can be deflected in the presence of applied current. 

\subsection{Generalized Snell's law}

Let us now examine the propagation of SWs when experiencing nonuniform current distributions. For simplicity, we start with the example of a SW propagating between two regions with different current densities $\textbf{j}$, \textit{i.e.} $\textbf{j}(x\leq0)=0$ and $\textbf{j}(x>0)=j_0\hat{y}$. It is well known that SWs reflect at material boundaries, where the momentum parallel to the interface should be conserved. In the case of SP current, the change in current density is equivalent to an interface, and the momentum perpendicular to $\nabla j$, \textit{i.e.}, $\textbf{k}\cdot\hat{\tau}\equiv |\textbf{k}-(\textbf{k}\cdot\hat{\nabla j})\hat{\nabla j}|$, with $\hat{\tau}$ the vector tangent to the interface, should be conserved. In our simple example that corresponds to $k^{(1)}_y=k^{(2)}_y$, where the indices 1 and 2 refer to the incident and refracted waves respectively. If the propagation direction is parallel to the $\textbf{k}$ vector, the well-known Snell's law applies: $k^{(1)}\sin(\phi_1)=k^{(2)}\sin(\phi_2)$, with $\phi_1$ and $\phi_2$ the incident and refracted angles respectively. However, since in our case the dispersion relation is asymmetric, the Snell's law has to be adjusted as follows
\begin{equation}\label{eq.Snell}
    k^{(1)}_g\sin(\phi_1)+\textbf{k}^{(1)}_{0}\cdot\hat{\tau}=k^{(2)}_g\sin(\phi_2)+\textbf{k}^{(2)}_{0}\cdot\hat{\tau},
\end{equation}
where the angles $\phi_i$ are taken with respect to $\hat{\nabla} j$ (\textit{i.e.}, the direction normal to the interface). Similarly generalized Snell's laws for the refraction of SWs at domain walls and heterochiral interfaces were derived in Refs.~\citenum{Yu2016} and \citenum{mulkers2018tunable}.   

\subsection{Micromagnetic simulations}

\begin{figure}[!t]
\centering
\includegraphics[width=0.6\linewidth]{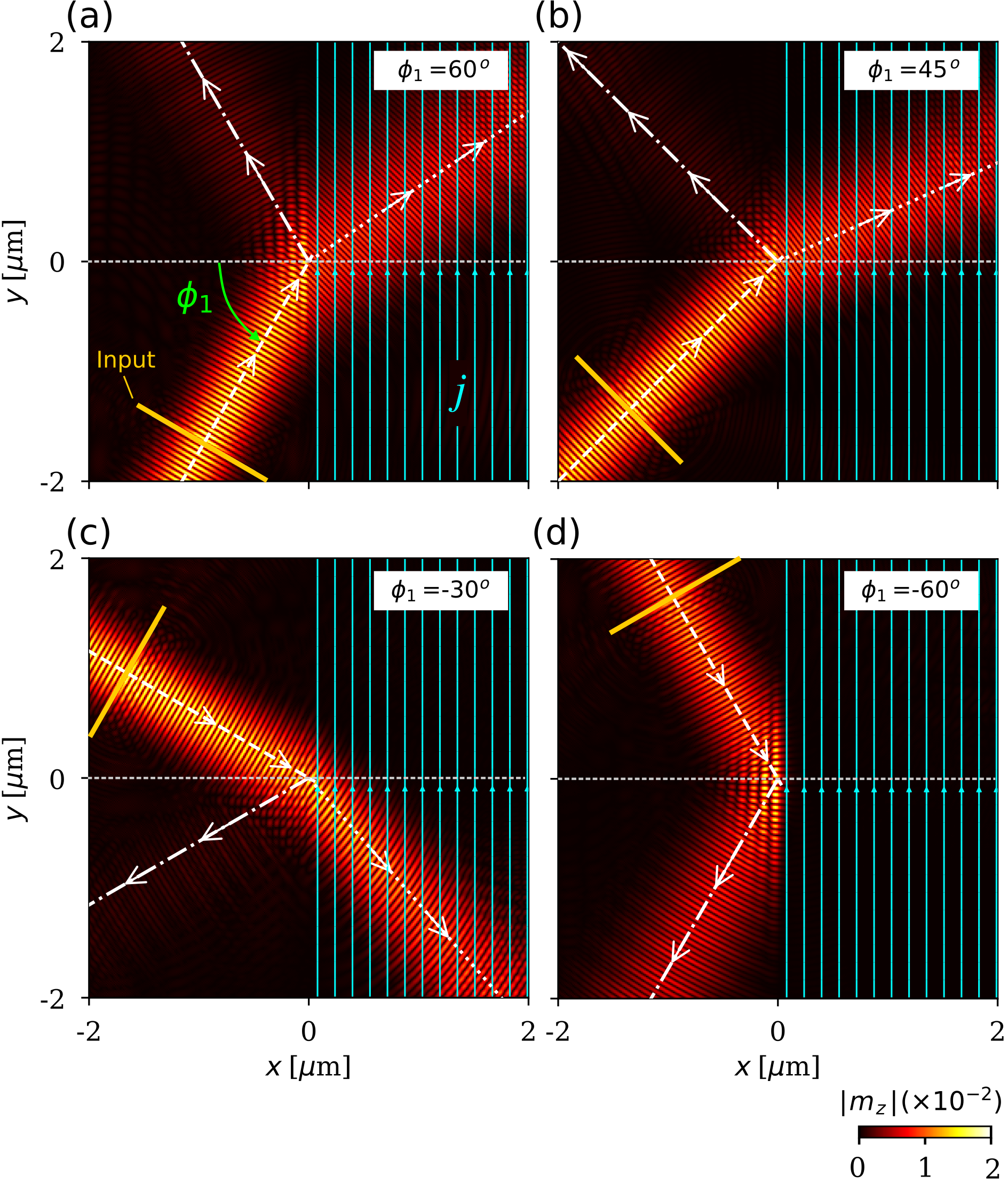}
\caption{\textbf{Generalized Snell's law for SW refraction}. (a-d) Snapshots of the simulated spin-wave propagation across a sharp interface (at $x=0$) where the current density changes from 0 to $\textbf{j}=j_0\hat{y}$, for different incident angles $\phi_1$. White arrows indicate the propagation direction following the generalized Snell's law [Eq.~\eqref{eq.Snell}]. In these simulations we considered SW frequency $f=20$~GHz, bias field $H_0=0.5$~T applied along $+\hat{x}$ direction, $\textbf{j}=2\times 10^{12}$~Am$^{-2}\hat{y}$, $\alpha=0.001$ and $\beta=0.002$.}
\label{fig2}
\end{figure}
\begin{figure}[!t]
\centering
\includegraphics[width=0.5\linewidth]{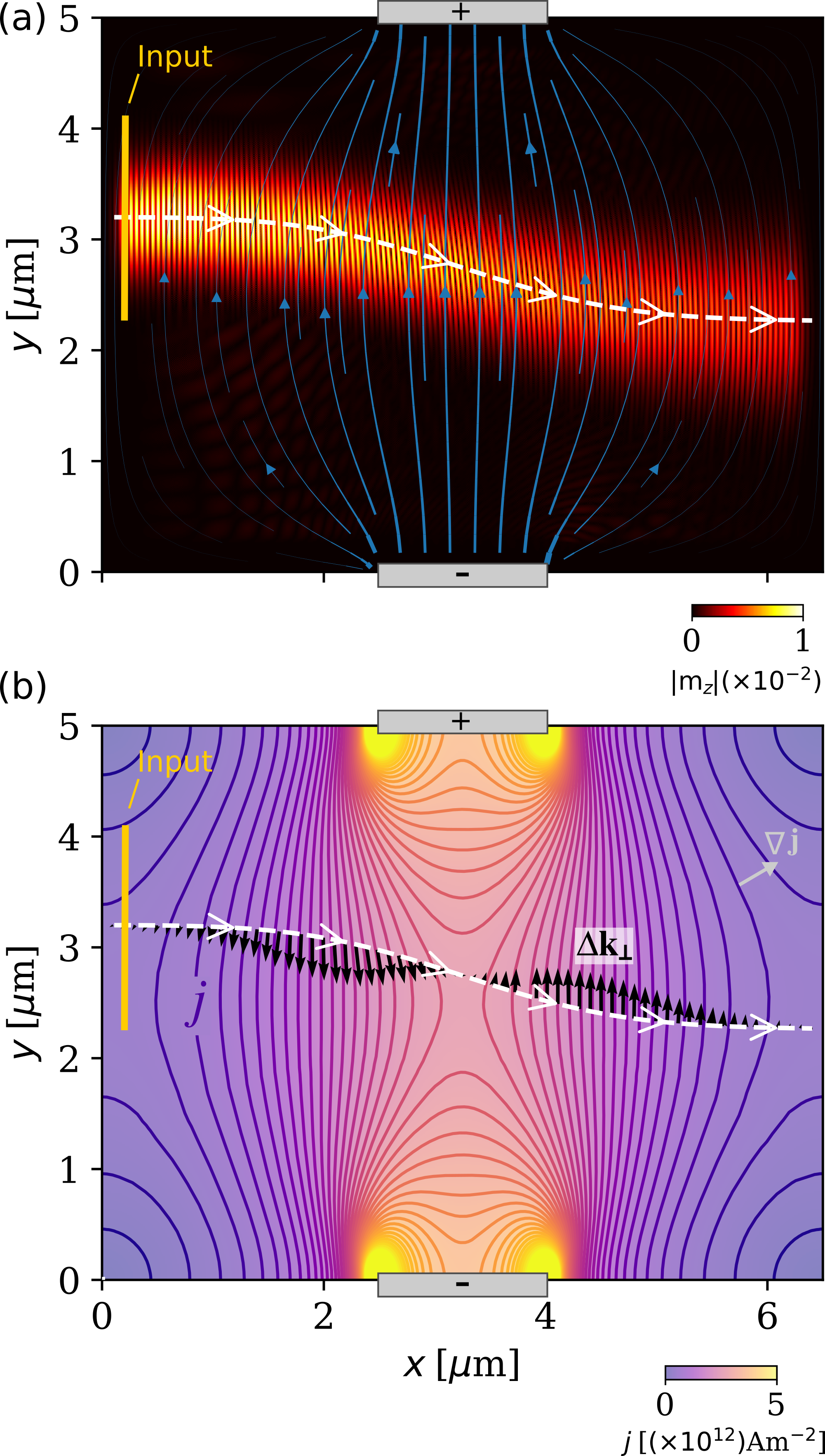}
\caption{\textbf{SWs under nonuniform current distribution}. (a) Snapshots of the SW propagating across a nonuniform current distribution induced between the voltage contacts (gray regions). White arrows indicate the SW trajectory calculated by iterating Eq.~\eqref{eq.Snell} locally along the current gradient in the propagation direction. Parameters are the same as in Fig.~\ref{fig2}. (b) Contour plot of the SP current density considered in (a). The shown isolines should be seen as interfaces where the SW is sequentially refracted during propagation. $\Delta \textbf{k}_\perp$ (black arrows) shows the accompanying change in the propagation direction (see text).}
\label{fig3}
\end{figure}

To realistically simulate a SW system in the presence of the SP current, we employ the micromagnetic framework and specifically its mumax$^3$ implementation~\cite{Vansteenkiste2014TheMuMax3,leliaert2018fast} to calculate the SW refraction when propagating between two regions with different current densities. 

The SW beams are created by a sinusoidal oscillating field $\textbf{h}=h_0\sin(\omega t)\hat{z}$ applied in a narrow rectangular region (input antenna, see Fig.~\ref{fig2}~(a)), where the field amplitude $h_0$ has a Gaussian profile in the transverse direction and $f=\omega/2\pi$ is the oscillation frequency. For all simulations we consider $h_0=0.01H_0$, \rmm{where $H_0=0.5$~T is the bias field applied along $+\hat{x}$ direction [see Methods section]}. Fig.~\ref{fig2}~(a-d) shows snapshots of the simulated SW propagation across the interface where the current density sharply changes, for different incident angles $\phi_1$. White arrows denote the SW trajectories predicted by our Eq.~\eqref{eq.Snell}, which are in excellent agreement with the micromagnetic simulations. Notice that the STT induced by the SP current can either deflect or confine the SWs. The critical incident angle $\phi^\ast$ 
above which the SW can not propagate for a given applied current, \textit{i.e.}, the SW undergoes total internal reflection, is obtained from our Snell's law by imposing that the refracted wave is parallel to the interface ($\phi_2=\pm\pi/2$), as
\begin{equation}\label{eq.phi_c}
    \phi^\ast=\pm\arcsin\left(\frac{\pm k^{(2)}_g +(\textbf{k}^{(2)}_{0}-\textbf{k}^{(1)}_{0})\cdot\hat{\tau}}{ k^{(1)}_g}\right).
\end{equation}
Note that due to the asymmetric dispersion relation, the refraction is not symmetric for positive and negative incident angles. This is seen in the results for $\phi_1=60^\circ$ and $\phi_1=-60^\circ$ in Fig.~\ref{fig2}~(a,d), where only for the second case the total reflection of the SW is achieved.   

~\\
\textit{\textbf{Poisson solver.}}---In practice, the current applied into the magnetic film will not exhibit a step-like distribution as considered in our previous example, but will spread continuously in the material obeying Poisson's equation~\cite{griffiths2005introduction}. To precisely calculate the interaction of SWs with such nonuniform current distributions, we implemented a Poisson solver in the micromagnetic simulation package mumax$^3$~\cite{Vansteenkiste2014TheMuMax3}. Fig.~\ref{fig3}~(a) shows a snapshot of the simulated SW propagation across the nonuniform SP current density created between shown finite voltage contacts at the sample edges. Notice that the SW is pertinently deflected while crossing the region where current is applied, to finally reach a shifted propagation direction upon leaving the area pierced by current. Although both the direction and magnitude of the current continuously change as the SW propagates, our generalized Snell's law still applies locally - the SW facing a local gradient in current density $\nabla j$ is effectively analogous to the earlier interface example. White arrows in Fig.~\ref{fig3}~(a) indicate the SW trajectory calculated by iterating Eq.~\eqref{eq.Snell} as the SW propagates in the current distribution, where in each step the angle $\phi_i$ is taken with respect to the local $\hat{\nabla} j$. The predicted trajectory is in very good agreement with the simulation, demonstrating the general applicability of Eq.~\eqref{eq.Snell} even for nonuniform current distributions. Fig.~\ref{fig3}~(b) shows the contour plot of the applied current density, where isolines can be seen as interfaces where the SW is refracted. $\Delta k_\perp=(\textbf{k}^{(2)}_{0}-\textbf{k}^{(1)}_{0})\cdot\hat{\tau}$ quantifies the change in the propagation direction of the SW, shown as black arrows in Fig.~\ref{fig3}~(b).

\subsection{Multichannel SW selector}

\begin{figure*}[!t]
\centering
\includegraphics[width=0.9\linewidth]{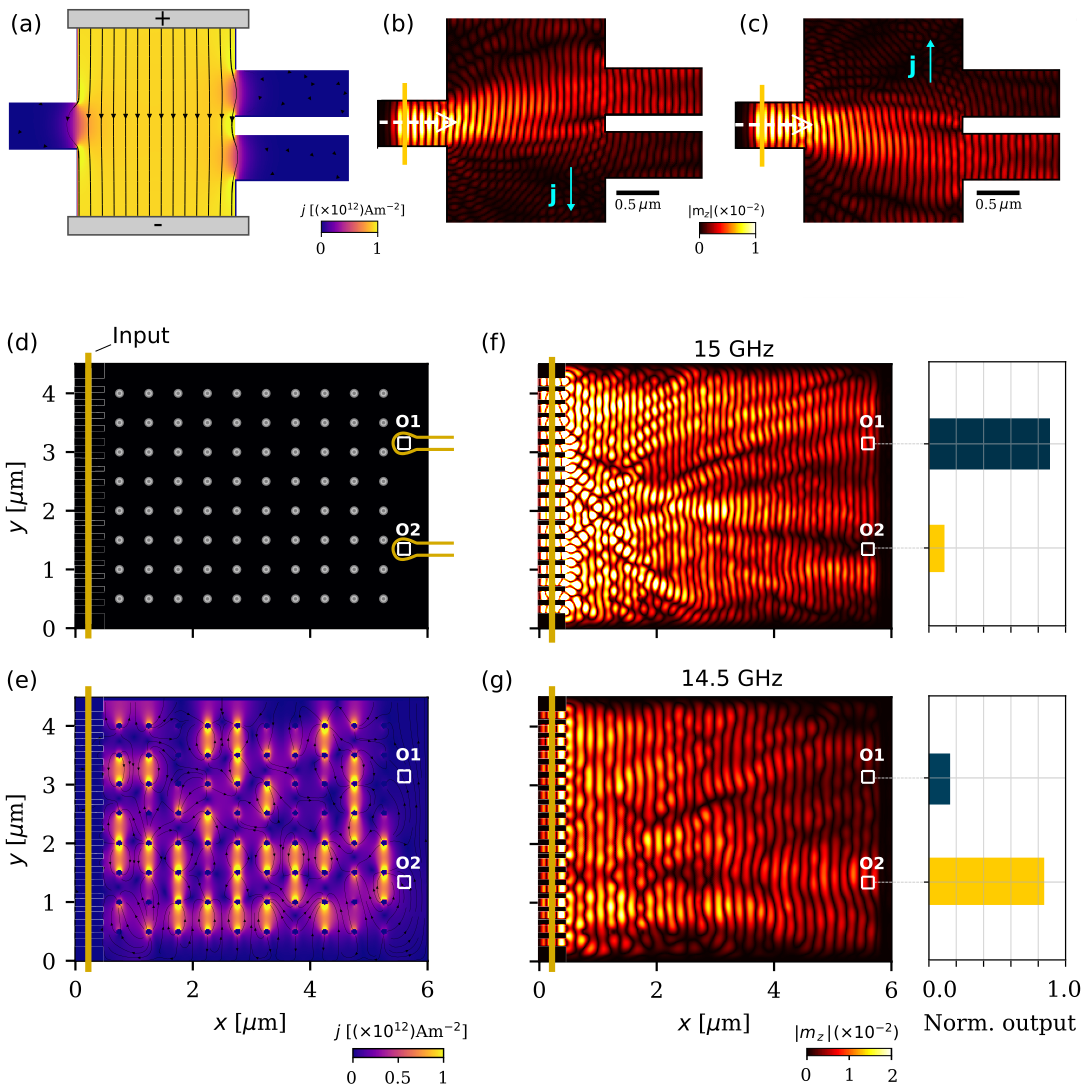}
\caption{\textbf{Guiding SWs for logic and neuromorphic computing.} (a) A multichannel SW selector demonstrated based on one input and two output channels. Voltage leads induce the SP current in the central region of the sample. (b-c) Snapshots of SW propagation simulated for the setup shown in (a). By changing magnitude or direction of the applied current one can guide the SW towards the desired output channel with minimal losses. (d) Scheme of the envisioned neural network hardware. The (18) input SWs are created on the left and propagate across the matrix of 80 voltage contacts (grey dots) of 100~nm in diameter each. (e) SP-current profile induced by the applied voltage that best performs the desired operation. (f-g) Snapshots of SW simulations after training the neural network. The voltage (current) pattern was trained to focus the waves of two different frequencies to the desired outputs. The bar charts show the normalized intensities at the output locations integrated over $10$~ns. In these simulations we considered SW frequency $f=15$~GHz (b, c, f) and $f=14.5$~GHz (g), and other parameters same as previously.}
\label{fig4}
\end{figure*}

Performing logic operations with SWs generally requires combining different input waves that interfere with each other to generate a desired logic output state~\cite{mahmoud2020introduction}. The ability to guide SWs through nanochannels is therefore vital to the development of more complex SW-based circuitry. In this regard, SP currents can be used to precisely guide the SW in such devices, for example, to selectively ``write" SWs in one of multiple nanotracks or logic gates in a larger microprocessor. We exemplify here such an application by simulating a multichannel SW selector, illustrated in Fig.~\ref{fig4}~(a). The input SWs are generated in a channel on the left-hand side of the sample and propagate across a region where SP current is applied. By tuning the magnitude or direction of the applied current one can precisely deflect the SWs towards one of the output channels on the right side of the sample, as shown in Figs.~\ref{fig4}~(b,c). Likewise, a frequency selector can be implemented considering the fact that SWs with different frequencies experience different deflections under the same applied current and can therefore be isolated into separate output channels.                 

\subsection{Neuromorphic computing}

Finally, we demonstrate the use of SP current for the design of neural-network hardware, based on SW propagation, where weights and interconnections of the network are realized by a pattern of the SP currents applied to the propagating substrate. Fig.~\ref{fig4}~(d) illustrates the envisioned device. The input signal is created on the left and propagates across a region with a matrix of 80 voltage contacts [gray dots in Fig.~\ref{fig4}~(d)] of 100~nm in diameter each. The read-out is taken from the two output antennas on the right side. An arbitrarily powered voltage matrix induces a distribution of the SP currents in the substrate that interacts with the input SWs. Training the neural network is equivalent to finding the current pattern that realizes the desired input-output mapping, for example, to classify different input signals by focusing them in different outputs. As suggested in Refs.~\citenum{hughes2019wave} and \citenum{papp2021nanoscale}, a back-propagation machine learning algorithm can be used for training a similar SW-based network, which can perform tasks such as vowel recognition and frequency classification. Here, we demonstrate that a simple Monte Carlo (MC) algorithm can perform the same task of training the neural network for simple classification problems. In our example, we perform a frequency-recognition operation, where we consider input SWs with frequencies \rmm{$f=15$ and $14.5$ GHz}. The neural network is trained to focus the SWs with \rmm{15~GHz} to the output O$1$ and SWs with \rmm{14.5~GHz} to the output O$2$ [see Figs.~\ref{fig4}~(f,g)]. The voltage at each contact is randomly initialized in one of the three values: $U_i=-u_0$, $0$ or $+u_0$, with \rmm{$u_0=0.15$~V}, and a Metropolis algorithm is implemented by changing the voltage of one of the contacts at every MC step [see supplemental material~\cite{SM}]. The finally trained configuration [resulting in the current distribution shown in Fig.~\ref{fig4}~(e)] is able to focus SWs of each frequency to the desired outputs as shown by the bar charts in Figs.~\ref{fig4}~(f,g), and can therefore perform the classification operation. \rmm{In the supplementary material, we show that the effects of Oersted and demagnetizing fields are negligible in our example~\cite{SM}. One should note however that the proposed neuromorphic application can be generalized for arbitrary scenarios, as long as all the relevant magnetic interactions are considered during the training process of the neural network. The here proposed fully electronic control enables facilitated reconfiguration of the neural network to perform different tasks within the same integrated circuit.}

\rmm{To identify the role of different material parameters in the current-induced SW scattering, and consequently in the proposed applications, let us consider the case of an SW propagating across two regions with different current densities, as in Fig.~\ref{fig2}. In this scenario, for an incident angle $\phi_1=0$ and making use of Eqs.~\eqref{eq.Snell} and \eqref{eq.u}, the refraction angle of the SW is given by $\phi_2=\sin^{-1}\left(\frac{ 1+\alpha\beta}{\sqrt{(1+\alpha\beta)^2 + \eta^2}}\right)$, where $\eta=\frac{4e\gamma}{\mu_B}\frac{Ak}{Pj}$. Therefore, the current-induced SW scattering is maximized when $\eta$ is minimized. In other words, the SW scattering is maximized in materials with small exchange stiffness $A$ and for SWs with small wavenumber $k$, as well as by increasing the SP current, represented by the polarization $P$ and applied current density $j$. Moreover, it is worth mentioning here that applied current $j$ may locally increase the temperature of the system, and consequently affect the magnetic parameters such as the saturation magnetization $M_s$, hence additional scattering may arise from such non-uniformity of the magnetic landscape in which SWs propagate. In Table~\ref{table_parameters} we show material properties of representative low-damping, metallic magnetic materials that can host spin waves and are therefore convenient candidates for the proposed control of SWs by SP current. As an additional example, we reproduced the neuromorphic application for the case of a Ni$_{80}$Fe$_{20}$ thin film in the supplementary material~\cite{SM}. }

\begin{table}[t!]
\begin{tabular}{lcccccccccc}
\hline\hline
  &$M_s$ (MA/m)    &$A$ (pJ/m)   & $\alpha$ ($\times10^{-3}$)          & $P$ $(\Omega m)^{-1}$              & References\\
\hline
Ni$_{80}$Fe$_{20}$ & 0.7 & 10 & 7 & 0.5  & [\citenum{Vlaminck2008,mahmoud2020introduction}]\\
CoFeB          & 1.3 & 15 & 4 & 0.65 & [\citenum{mahmoud2020introduction,huang2008spin}]\\
CoFeAlB        & 1.0 & 9  & 3 &      & [\citenum{conca2017cofealb}]\\
\hline\hline 
\end{tabular}
\caption{Material properties of representative low-damping, metallic magnetic materials that can host spin waves and are therefore candidates for the proposed control of SWs by SP current.}  
\label{table_parameters}
\end{table}

\section{Conclusions}

We have demonstrated the use of non-uniform spin-polarized current for the manipulation of spin-waves. We showed that the spin-transfer torque induced by the applied current has an effect analogous to DMI for confining spin-waves and controlling their propagation direction in the linear regime, and derived a generalized Snell's law that describes the scattering of spin-waves between regions with different current densities. Finally, we implemented the calculation of the current distribution in micromagnetic simulations by solving the Poisson's equation within the simulation package mumax$^3$ (to be made available in the upcoming release of mumax$^4$), in order to (i) validate the derived Snell's law and (ii) demonstrate how strategically applied current distributions can be employed in magnonic logic and neuromorphic computing devices, thereby advancing the prospects of low-power spintronic hardware and artificial intelligence.

\section{Methods}

\subsection{Micromagnetic simulations}
For the simulations, we employ the micromagnetic framework MUMAX$^3$~\cite{Vansteenkiste2014TheMuMax3,leliaert2018fast}, where we consider $10$~nm thick magnetic films with saturation magnetization $M_\text{s}=0.14$~MAm$^{-1}$, exchange stiffness $A=3.65$~pJm$^{-1}$~\cite{papp2017nanoscale} and damping constant $\alpha=0.001$. 
The considered free energy density is given by Eq.~\eqref{eq.E}, and the dynamics of the magnetization is governed by the LLG equation [Eq.~\eqref{eq.llg}] with STT terms [Eq.~\eqref{eq.stt}] accounting for in-plane applied currents. The in-plane bias field is set to $\textbf{H}=H_0\hat{x}$, with $H_0=0.5$~T. For all simulations, we consider a system discretized into cells of size $10\times10\times10$ nm$^3$. The polarization rate of the SP current is fixed at $P=0.4$. The calculation of the current distribution is made by solving the Poisson’s equation for the considered geometries, where we assume the electrical conductivity as $\sigma=1\times10^6$~($\Omega$m)$^{-1}$. 

\begin{acknowledgement}

This work was supported by the Research Foundation - Flanders (FWO-Vlaanderen, under Grant No. K226322N and 12A9223N), EoS ShapeME project, and Brazilian Agencies FACEPE, CAPES and CNPq.

\end{acknowledgement}

\begin{suppinfo}

Calculation of spin-wave dispersion relation under spin-polarized current; Monte Carlo training of neural network; Effect of Oersted and demagnetizing fields; Example of a spin-wave neural network in a Ni$_{80}$Fe$_{20}$ thin film.

\end{suppinfo}

\bibliography{references}

\end{document}